\documentclass[reprint]{JASA}

\usepackage{siunitx}
\newcommand{\RNum}[1]{\uppercase\expandafter{\romannumeral #1\relax}}

\usepackage{tikz}
\newcommand\copyrighttext{%
  \footnotesize Copyright 2026 Acoustical Society of America. This article may be downloaded for personal use only. Any other use requires prior permission of the author and the Acoustical Society of America.}
\newcommand\copyrightnotice{%
\begin{tikzpicture}[remember picture,overlay]
\node[anchor=south,yshift=10pt] at (current page.south) {\fbox{\parbox{\dimexpr\textwidth-\fboxsep-\fboxrule\relax}{\copyrighttext}}};
\end{tikzpicture}%
}

\begin{document}

\title[]{Room compensation for loudspeaker reproduction using a supporting source}

\DOInumber{10.1121/10.0043238}

\author{James Brooks-Park}
\affiliation{Acoustics Group and Cluster of Excellence "Hearing4all", Carl von Ossietzky Universität Oldenburg, Oldenburg, Germany}

\author{Søren Bech}
\affiliation{B\&O Research, Bang \& Olufsen A/S, Struer, Denmark}
\affiliation{Department of Electronic Systems, Aalborg University, Aalborg, Denmark}

\author{Jan Østergaard}
\affiliation{Department of Electronic Systems, Aalborg University, Aalborg, Denmark}

\author{Steven van de Par}
\affiliation{Acoustics Group and Cluster of Excellence "Hearing4all", Carl von Ossietzky Universität Oldenburg, Oldenburg, Germany}

\email{James.Brooks-Park@Uni-Oldenburg.de}

\begin{abstract}
Room compensation aims to improve the accuracy of loudspeaker reproduction in reverberant environments. Traditional methods, however, are limited to improving only spectral (timbral) and temporal accuracy, neglecting the spatial accuracy of loudspeaker reproduction. Proposed is a method that compensates for both spectral and spatial properties of loudspeaker reproduction, by adding energy to the perceived reverberant sound field in a frequency-selective manner using a delayed secondary supporting source. This approach allows for the modification of the Direct to Reverberant Ratio (DRR) as a function of frequency, altering spatial and spectral reproduction. The proposed method is perceptually evaluated, demonstrating its ability to alter the perception of a primary loudspeaker without the listener perceiving the supporting source. The results show that the proposed method performs comparably to a well-established commercial room compensation algorithm, and has several advantages over traditional room compensation methods.
\end{abstract}

\maketitle
\copyrightnotice

\section*{Introduction}
The accuracy of loudspeaker reproduction in reverberant environments is determined by a combination of the free-field response of the loudspeaker\footnote{The anechoic response of the loudspeaker across all azimuth and elevation angles.}, the acoustical and geometric properties of the playback room, and source and receiver positions. Whilst the free-field response of the loudspeaker can be optimised by the manufacturer, the reverberant sound field is largely uncontrolled. Room equalisation, or compensation, is often employed to address artefacts introduced by the reverberant sound field, increasing playback accuracy, with the aim of room-independent loudspeaker reproduction.

In reverberant environments, each source and receiver position results in a unique Loudspeaker-Room Impulse Response (LRIR). Loudspeaker directivity determines the frequency-dependent distribution of energy between the direct and reverberant sound fields, influencing the spectrum of the Loudspeaker-Room Transfer Function (LRTF). The frequency-dependent nature of loudspeaker directivity is a consequence of the relationship between the reproduced frequency and the driver and cabinet size. A conventional two-way loudspeaker reproduces lower frequencies in a more omnidirectional pattern. However, as frequency increases, the wavelength decreases relative to the size of the driver, forming a more directional on-axis beam, resulting in more low-frequency energy being emitted into the room compared to higher frequencies.

The acoustical characteristics of the playback room also impact loudspeaker reproduction. The shape and dimensions of the playback room dominate LRTF characteristics below the Schroeder (transition) frequency (\SI{200}{\Hz}-\SI{300}{\Hz}) \cite{schroeder_frequency_1962}, determining resonant frequencies. Above this frequency, the LRTF becomes stochastic, where absorption coefficients of reflective boundaries further shape the LRTF.

Whilst the majority of research into auditory perception of acoustic spaces has been focused on concert halls and other architectural acoustic spaces \cite{beranek2008concert}, the effect of ordinary-sized listening rooms on loudspeaker reproduction has been investigated \cite{toole2017sound}. \citet{kaplanis2019perception} shows that the acoustic space, independent of the loudspeaker, significantly affects the perceived auditory experience. The main effects found to affect perceptual differences were \textit{reverberance}, \textit{width and envelopment}, and \textit{proximity}.


Spatial and distance perception should also be considered in loudspeaker reproduction. One underlying mechanism that facilitates acoustical distance perception is the Direct to Reverberant Ratio (DRR) \cite{zahorik2002direct}. This ratio compares the energy in the direct sound, propagating directly from the source to the receiver, to the energy in the reverberant sound, reaching the receiver only after one or more reflections. Disregarding visual or loudness cues, in acoustic environments, humans rely on the DRR for source distance perception; variations to the DRR across frequencies and position may alter the spatial image of the source. 

When the on-axis free-field amplitude response of a loudspeaker is frequency-independent, the frequency-dependent directivity pattern, in combination with the acoustical properties of the room, shape the LRTF spectrum. In these cases, the frequency-dependent LRTF energy is coupled to the frequency-dependent DRR, linking spectral and spatial (distance) perception.

The control of the DRR can significantly impact immersive playback by accurately positioning sound sources. \citet{laitinen2015controlling} demonstrate that controlling source directivity, thereby altering the DRR, can influence distance perception, showing a greater impact than only changing the loudness of the source. However, whilst the DRR across the audible frequency range could be controlled in this manner, spatial lobes in the directivity pattern at higher frequencies restricted control.

Over the past 50 years, numerous techniques have been proposed in the field of room compensation (equalisation), aiming to compensate for inaccuracies introduced by loudspeakers and playback rooms (See \cite{cecchi_room_2017} for a review). Traditionally, for stationary single-channel systems, room compensation techniques aim to design a filter based on a known LRIR, which, when applied to the loudspeaker, compensates for spectral irregularities in the sound field at the receiver position. However, to ensure such methods work in practice, several considerations must be made, increasing the complexity of the problem.

Due to the stochastic nature of the reverberant sound field, the resulting filter from a room compensation algorithm can be expected to be comparably complex, and may introduce perceptual artefacts \cite{neely_invertibility_1979}. Inverting notches (zeros) in the LRIR can introduce poles in the inverse filter, potentially leading to audible resonances or instability. Pre-ringing or pre-echo filter artefacts are also a concern for non-minimum phase LRIRs, potentially degrading reproduction accuracy, as opposed to remedying.

Due to the variability of LRIRs within a listening area, a filter designed for one position is not guaranteed to be valid for any other position, or indeed at any other time than the time of measurement. In some cases, using a filter designed for a different position can reduce reproduction accuracy \cite{mourjopoulos_variation_1985}. For this reason, room compensation methods are typically only effective at low frequencies, where positional variation is minimal.

Several techniques have been proposed to address the discussed issues, aiming to design filters that avoid introducing perceptual artefacts and improve spatial robustness. A common approach to increase spatial robustness is to design a filter based on several spatially distributed LRIRs. \citet{elliott_multiple-point_1989} demonstrates that incorporating several LRIRs in the filter design, rather than a single LRIR, improves the spatial robustness of the resulting filter. More recently, machine learning approaches have also been proposed to create an average room transfer function based on a single LRIR measurement \cite{tuna2022data}. These approaches minimise the error across several positions, preventing overfitting to a single LRIR. Building on the concept of distributed measurements, \citet{brannmark_robust_2008} explore how zeros in the transfer function vary across measurement positions, posing a challenge to filter design at these frequencies. A proposed method involves inverting only the transfer function at frequencies with limited variation. Whilst this approach does not compensate for all frequencies, the risk of pre-ringing is reduced.
    
Regularisation is another technique of increasing robustness and minimising perceptual artefacts. Reducing the total energy in a filter, large peaks, notches, and variations are not carried over from the LRIR to the filter \cite{kirkeby_fast_1998}. A similar concept to regularisation, smoothing, is often used in magnitude-based room compensation algorithms. Smoothing modifies the LRTF by reducing abrupt spectral features, ensuring features that are likely to vary from position to position are not carried across to the resulting compensation filter. Specifically, octave-based smoothing, including higher-resolution subdivisions such as 1/3-octave, is used for the purpose of room compensation \cite{hatziantoniou_generalized_2000}, whereby the size of the smoothing window increases with centre frequency, in line with the sensitivity of the human auditory system.

Room compensation literature often overlooks compensation for spatial perception particularly distance perception. As discussed, the DRR is partially responsible for acoustic distance perception, and is a product of loudspeaker directivity and the acoustical properties of the playback room. In playback rooms with a variable $T_{60}$ time across frequencies, or for loudspeakers with directivity characteristics that vary with frequency, the DRR will change as a function of frequency and the acoustical distance perception will be frequency-dependent. Traditional room compensation techniques apply a filter to the free-field response of the loudspeaker, altering the spectral response of both the direct and the reverberant sound fields. Unless full deconvolution is achieved (the difficulties of which are discussed in \cite{mourjopoulos_variation_1985}), the DRR is left uncontrolled.

First proposed in \cite{brooks2023reverberant}, this publication investigates a method of room compensation, whereby the playback room and loudspeaker directivity are compensated for by altering the perceived reverberant sound field. The proposed method aims to alter the DRR as a function of frequency, compensating for both spectral and spatial inaccuracy of loudspeaker reproduction in reverberant environments.

By introducing an additional source (a supporting loudspeaker to a primary loudspeaker), energy can be added to the room in a frequency-selective manner, compensating for spectral notches in the primary source's spectrum. Delaying the supporting source by an appropriate $\Delta (t)$, activates the precedence effect, ensuring only the primary loudspeaker is spatially perceived as a source.

The paper is structured as follows: Section \ref{sec:ResearchQuestion} defines the research question, Section \ref{sec:Theory} investigates how traditional room equalisation differs to the proposed approach, defining the theoretical basis behind the proposed method. Section \ref{sec:Implementation} presents an implementation of the proposed reverberant sound field compensation method; Section \ref{sec:PE} subjectively analyses the performance of the proposed approach compared to traditional room compensation, Section \ref{sec:ObjMes} technically investigates the proposed method. Section \ref{sec:DRRAnalysis} analyses how, in practice, the proposed approach is able to alter the DRR of the sound field, Section \ref{sec:GeneralDiscussion} discusses the proposed approach and the presented evaluations. Section \ref{sec:Conclusion} concludes the paper, stating the main contributions.

\section{Research Question}\label{sec:ResearchQuestion}
The aim of this publication is to investigate whether the perception, both spatial and timbral, of a primary loudspeaker can be altered by a supporting compensation loudspeaker, without the supporting loudspeaker being spatially perceived as an additional source. To validate this, a perceptual evaluation is conducted investigating the proposed method and forming a comparison against a well-established commercial room compensation algorithm.

\section{Room Compensation}
\label{sec:Theory}
\subsection{Transfer function representation}
In an acoustic environment, the LRTF from source to receiver can be expressed as the sum of the direct and the reverberant sound transfer functions as a function of angular frequency ($\omega$), 
\begin{equation}
    h(\omega) = h^{\text{dir}}(\omega) + h^{\text{rev}}(\omega) \; . \label{eq:LRTFRepresentation}
\end{equation}

The direct sound, $h^{\text{dir}}$, is modeled as a delta pulse centered at the arrival time of the first wavefront $t = n_1$, where the peak of the delta pulse occurs:
\begin{equation}
     h^{\text{dir}}(t) = 
    \begin{cases}
        h(t), & t = n_1 \\
        0, & t \neq n_1
    \end{cases}
    \; .
\end{equation}

The reverberant sound field transfer function, $h^{\text{rev}}$, represents the portion of the impulse response that arrives at the receiver after the direct sound: 
\begin{equation}
    h^{\text{rev}}(t) = 
    \begin{cases}
        0, & t \leq n_1 \\
        h(t), & t > n_1
    \end{cases}
    \; .
\end{equation}

Figure \ref{fig:dir,er,la} shows the separation of the direct sound, early, and late reflections. As per our definition, the reverberant sound is a combination of the early and late reflections. 

\begin{figure}
    \begin{center}
        \includegraphics[width=\linewidth]{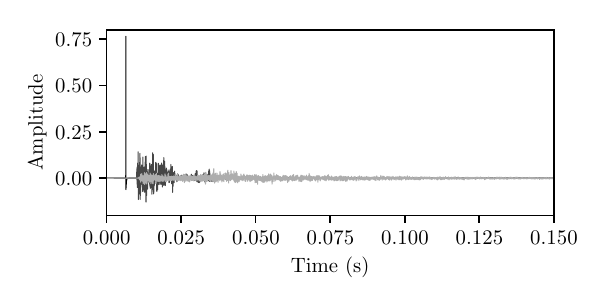}
         \caption{\label{fig:dir,er,la}Room impulse response generated by RAZR \cite{wendt_computationally-efficient_2014} depicting the discretisation of the direct sound (black), early reflections (dark grey), and late reflections (light grey).}
    \end{center}
\end{figure}

Loudspeaker directivity patterns, uneven boundary absorption coefficients, and frequency-dependent reverberation times contribute to an uncontrolled distribution of energy across the audible frequency range in the power spectrum,$|h(\omega)|^2$, of the LRTF. This not only affects the spectral reproduction accuracy but also the DRR as a function of frequency, compromising spatial reproduction accuracy,
\begin{equation}\label{eq:DRR}
    DRR_{\text{h}}(\omega) = \frac{|h^{\text{dir}}(\omega)|^2 }{ |h^{\text{rev}}(\omega)|^2} \; .
\end{equation}
Where $DRR_{h}(\omega)$ represents the DRR of the sound field $h$.

\subsection{Traditional Compensation}
\label{sec:TradMethod}
Many room compensation methods have been proposed with the aim of compensating for the uncontrolled energy distribution in the LRTF by applying a frequency-dependent filter $w(\omega)$, to the free-field response of the loudspeaker. As shown in Figure \ref{fig:TradDiagram}, represented as,
\begin{equation}
    h_{\text{opt}}(\omega) = w(\omega) \left(h^{\text{dir}}(\omega) + h^{\text{rev}}(\omega) \right) \;.
\end{equation}

Whilst this approach is able to control the spectral energy distribution of the LRTF, the DRR is left uncontrolled (unless full dereverberation is achieved),
\begin{equation} \label{eq:TradOptDRR}
    DRR_{\text{h,opt}}(\omega) = \frac{|w(\omega)h^{\text{dir}}(\omega)|^2 }{ |w(\omega)h^{\text{rev}}(\omega)|^2} \;.
\end{equation}

\subsection{Proposed Method}
\label{sec:PropMethod}
\begin{figure}
\centering
    \begin{minipage}{0.5\textwidth}
        \includegraphics{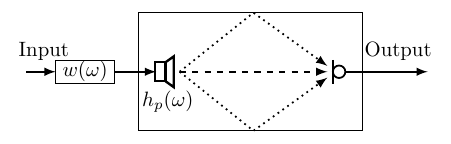}
    \end{minipage}
    \caption{\label{fig:TradDiagram}Traditional Room compensation ideology comprising of the LRTF $h_p(\omega)$, and a compensation filter $w(\omega)$}
    \begin{minipage}{0.5\textwidth}
        \includegraphics{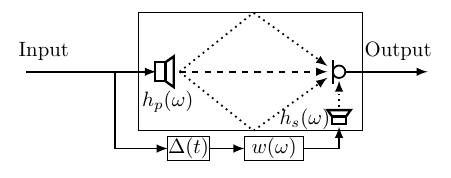}
    \end{minipage}
    \caption{\label{fig:PropDiagram}Proposed room compensation ideology. Introducing the supporting source $h_s(\omega)$, the supporting compensation filter $w(\omega)$, and the delay $\Delta (t)$, to compensate for the primary loudspeaker LRTF $h_p(\omega)$}
\end{figure}

The proposed method of room compensation introduces a secondary supporting source with the aim of altering the perceived reverberant sound field. By appropriately delaying ($\Delta (t)$) and modifying the LRTF of the supporting source, energy can be added to the reverberant sound field in a frequency-dependent manner (as shown in Figure \ref{fig:PropDiagram}). This is achieved without the listener perceiving an additional source, as the precedence effect is activated, instead perceiving additional reverberation,
\begin{equation}
    h_{\text{opt}}(\omega) = h^{\text{dir}}_p(\omega) + h^{\text{rev}}_p(\omega)+ 
                w(\omega)\left(h^{\text{dir}}_s(\omega) + h^{\text{rev}}_s(\omega)\right) \; .
\end{equation}
Where the transfer function of the primary source is $h_p(\omega)$ and the supporting source $h_s(\omega)$.

As previously defined, the reverberant sound field comprises of all energy arriving at the receiver position after the direct component, allowing the formulation of the proposed method to be simplified,
\begin{equation}
    h_{\text{opt}}(\omega) = h^{\text{dir}}_p(\omega) + h^{\text{rev}}_p(\omega)+ 
                w(\omega)h'^{\text{ rev}}_s(\omega) \;,
\end{equation}
where $h'^{\text{ rev}}_s(\omega)$ indicates the entirety of the delayed supporting source, categorised as reverberant sound based upon its relative position in time to the direct sound of the primary loudspeaker.

As the frequency-dependent filter $w(\omega)$ is applied to the supporting source, modifying the perceived reverberant sound field, both the power spectrum, $|h_{\text{opt}}(\omega)|^2$, and the DRR can be modified as a function of frequency,
\begin{equation}\label{eq:PropOptDRR}
    DRR_{\text{h,opt}}(\omega) = \frac{|h^{\text{dir}}_p(\omega)|^2 }{|h^{\text{rev}}_p(\omega)+ 
    w(\omega)h'^{\text{ rev}}_s(\omega)|^2} \;,
\end{equation}
compensating both for spectral and distance perception.

\section{Proposed Implementation}\label{sec:Implementation}
The proposed method aims to compensate for both spectral and spatial inaccuracies introduced by loudspeaker directivity and playback room acoustics by altering the perceived reverberant sound field. To alter the perceived reverberant sound field independently of the direct sound, a supporting loudspeaker is introduced, spectrally compensating for the primary loudspeaker. The supporting loudspeaker reproduces a filtered version of the content reproduced by the primary loudspeaker, filling in spectral notches in the primary LRTF. The signal path of the supporting loudspeaker must be delayed, ensuring the supporting loudspeaker is not spatially perceived as an additional source.

Assuming the on-axis free-field response of a loudspeaker to have a frequency-independent amplitude response, inconsistencies in the magnitude response of the LRIR are a result of the reverberant sound field. By including this assumption in our filter design, the spectrum of the LRIR can be used to design compensation filters, as the direct sound is assumed to not alter the magnitude response of the LRTF.

\subsection{Filter Design}
The proposed method aims to add energy to the perceived reverberant sound field, filling in spectral notches in the LRTF of the primary loudspeaker and thus relies on the summation of energy at the receiver position. The supporting source filter $w(\omega)$ is to be designed such that when applied to the supporting loudspeaker $h_s(\omega)$,  the combined spectral transfer function with the primary loudspeaker $h_p(\omega)$ at the receiver position matches the target function $|d(\omega)|$,
\begin{equation}
    |d(\omega)| = \sqrt{|h_p(\omega)|^2 + |w(\omega) h_s(\omega)|^2} \;.
\end{equation}

The filter is calculated by inverting the difference in energy between the  target function and the transfer function of the primary loudspeaker, over the magnitude response of the supporting loudspeaker,
\begin{equation} \label{eq:ProposedComp}
    w(\omega) = \frac{\sqrt{|d(\omega)|^2-|h_p(\omega)|^2}}{|h_s(\omega)|} \;.
\end{equation}
In order for this design to be valid, it should be ensured that for each frequency, the magnitude of the target function is equal to, or greater than the magnitude of the primary loudspeaker, in order to avoid negative values in the inversion. Steps to ensure this are described further in Section \ref{sec:targetfunctiontechnical}.

This frequency domain filter can subsequently be converted into a time domain filter via an IFFT, and converted to minimum phase to remove any cyclic time domain properties \cite{smithfilters}. As the filter is applied to the reverberant sound field, the length of the filter can be longer than that of a traditional room compensation filter. A filter length of $2^{13}$, which with a sampling frequency of \SI{44.1}{\kHz} has a length of \SI{0.186}{\s}, has been selected to have suitable control over the frequency range subject to compensation.

\subsection{Delay}
\begin{figure}
    \begin{center}
        \includegraphics[width = \linewidth]{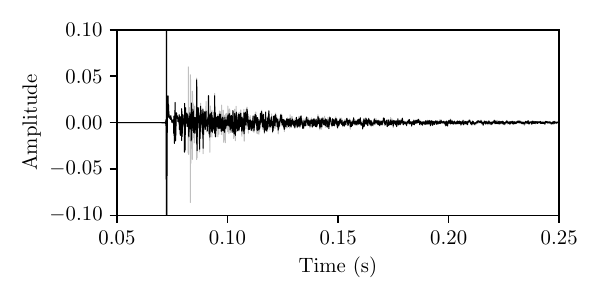}
        \caption{\label{fig:TimeDomainCompensation} Two time domain LRIRs are presented, one with (grey) and one without (black) reverberant sound field compensation.}
    \end{center}
\end{figure}

To spatially conceal the supporting loudspeaker in the perceived reverberant sound field, the precedence effect is utilised, ensuring the listener does not perceive the supporting loudspeaker as an additional source. By delaying the supporting source by 2-50ms, the listener will only spatially perceive the primary loudspeaker. \citet{walther_effect_2013} show that the content of the lagging source does not need to be identical to the leading source, allowing the spectrum of the supporting source to differ from the primary source. By delaying the supporting loudspeaker by a $\Delta (t)$, the physical characteristics of the reverberant sound field are also altered, as the precedence delay represents the onset delay of the reverberant sound field. This can be seen in Figure \ref{fig:TimeDomainCompensation}, demonstrating that the direct sound is left unaltered, whilst energy is added to the reverberant sound field.

For the presented implementation, a \SI{10}{\ms} delay is used; this was found by the authors to activate the precedence effect without significantly increasing the perceived $T_{60}$ time. The \SI{10}{\ms} delay has been chosen to be large enough to activate the precedence effect, large enough to be spatially robust if the listener were to move away from the sweet spot and closer to the supporting source, but small enough not to introduce echoes. \SI{10}{\ms} was found to be suitable across several rooms used to develop the proposed method. The calculation of the exact delay applied to the supporting loudspeaker takes into account the transmission delay of both the primary and supporting loudspeakers, ensuring the supporting loudspeaker is delayed by exactly \SI{10}{\ms} at the listening position relative to the primary loudspeaker.

\subsection{Loudspeaker Interaction}
In order to ensure purely energetic addition of the two sources, the supporting loudspeaker is decorrelated using a sparse noise sequence. The sparse noise sequence `velvet noise' \cite{alary2017velvet}, has been selected to decorrelate the supporting loudspeaker, altering the phase response of the loudspeaker whilst maintaining the desired magnitude response. In addition to the supporting loudspeaker delay, these methods allow the loudspeakers to act independently, ensuring the listener's auditory system processes the sum of the energy from both sources.

\subsection{Impulse response conditioning}
One of the primary issues discussed in the Introduction is the difficulty in designing a compensation filter that is able to compensate for more than a single position in a listening area. In the proposed implementation, several methods are employed to create a spatially robust filter. By calculating the average power response across two receiver positions within the listening area, the effect of spectral features that are not present across all receiver positions is reduced. Receivers are positioned 17cm apart to match the distance between the two ears. In order to uniformly measure the primary and supporting loudspeakers in the same receiver position, a microphone positioned facing upwards ensures the microphone response is symmetrical for both sources. The second conditioning technique is spectral smoothing, where the LRTF magnitude response is smoothed removing sharp notches and peaks that may result in a spatially non-robust filter. Given the relatively small distance between microphone positions, magnitude smoothing aims to increase the size of the listening area by removing abrupt changes to the LRTF that may not be consistent across the entire listening area. 1/3 Octave-based magnitude smoothing is used to have a smoothing window that changes as a function of frequency, with a similar frequency resolution to the human auditory system.

\subsection{Target Function (Technical)}\label{sec:targetfunctiontechnical}
\begin{figure}
    \begin{center}
        \includegraphics[width=\linewidth]{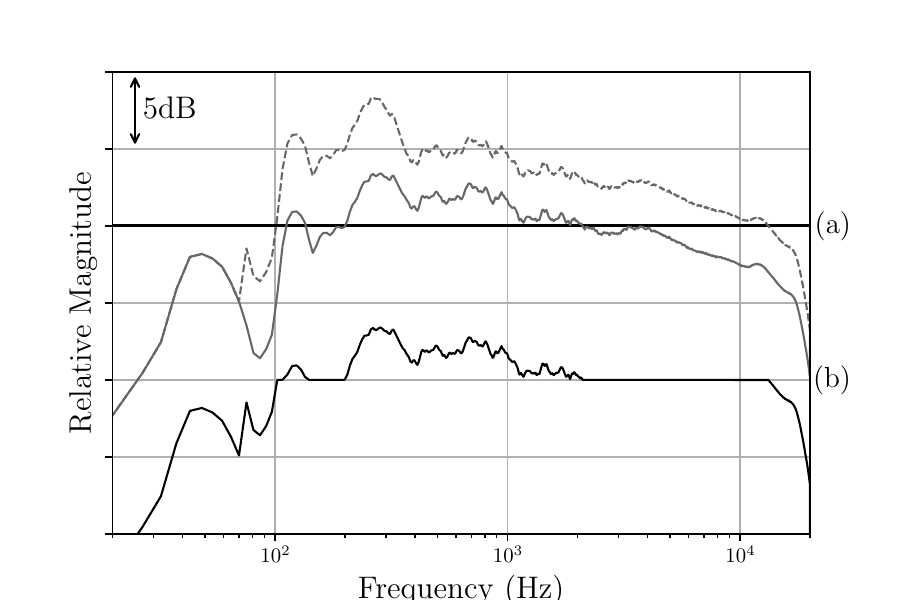}
        \caption{\label{fig:targetlimitations}This example illustrates the limitations introduced to the target function to maintain the precedence effect and prevent the resulting filter from removing energy from the room. (a) shows the original target function $d$ (black), the LRTF $h_p$ (grey), and the LRTF limit $h_{p,lim}$ (grey, dashed). (b) depicts the resulting target function,$d_{mod}$, after limiting.}
    \end{center}
\end{figure}

The desired spectral target function $d(\omega)$ must be modified such that two filter constraints are fulfilled (Visually demonstrated in Figure \ref{fig:targetlimitations}). The supporting loudspeaker should be limited such that energy is only added to the perceived reverberant sound field, and not attempt to cancel energy emitted by the primary loudspeaker. In order to restrict the filter from energetic cancellation, the target must be modified such that the frequency-dependent amplitude of $d(\omega)$ is greater than or equal to $h_p(\omega)$,
\begin{equation}\label{eq:Lim1}
    d_{mod}(\omega) = \left\{ \begin{array}{rcl}
        {d(\omega)} & \mbox{for}
        & d(\omega) \geq h_p(\omega) \\ h_p(\omega) & \mbox{for} & h_p(\omega) > d(\omega) \\
        \end{array}\right. \;.
\end{equation}

The amplitude of the supporting loudspeaker must be limited per frequency in order to not break the precedence effect. For a certain lead-lag time, the precedence effect will be broken if energy of the lagging source exceeds the energy of the leading source by a given threshold $\mathcal{T}$\,dB. \citet{walther_effect_2013} show that as frequency increases, the threshold at which the precedence effect is broken decreases, therefore the threshold can be expressed as a function of frequency $\mathcal{T}(\omega)$. The implication of this limit is a target function that does not exceed $ h_{p,lim}(\omega) = 10^{(20log(|h_p(\omega)|) + \mathcal{T}(\omega))/20}$,
\begin{equation}\label{eq:Lim2}
    d_{mod}(\omega) = \left\{ \begin{array}{rcl} 
        d(\omega) & \mbox{for}& h_{p,lim}(\omega)\geq d(\omega) \\
        h_{p,lim}(\omega)& \mbox{for} & d(\omega) > h_{p,lim}(\omega)  \\
        \end{array}\right. \;.
\end{equation}
Chosen by the authors to ensure the supporting source is not perceived, between the frequencies \SI{70}{\Hz} and \SI{500}{\Hz}, a precedence limit of \SI{10}{\dB} has been selected and between \SI{500}{\Hz} and \SI{20}{\kHz}, a limit of \SI{6}{\dB}. Figure \ref{fig:targetlimitations} (c) demonstrates how the two constraints affect the spectrum of the target function.

With the described target function constraints, the amplitude of the unmodified target function can be optimised to reduce the amount to which the constraints alter the target function. The amplitude of the target function can be set to increase/decrease the spectral peaks or precedence ensuring notches that remain in the modified target function. For the proposed implementation, the amplitude has been optimised to have an equal deficit, in decibels within the frequency region \SI{70}{\Hz} to \SI{10}{\kHz}, between the deviations to the target function caused by both constraints.

Also implemented into the modified target function is bandwidth limitation. In order to ensure only a band limited frequency range is subject to compensation, frequencies outside of this range (\SI{70}{\Hz} to \SI{20}{\kHz}) have the same target amplitude as the unmodified primary LRTF.

\subsection{Target Function (Subjective)}
Whist the precise target function for a loudspeaker reproduction may be a subjective choice, the chosen magnitude target aims to resemble the power spectrum of a loudspeaker with conventional two-way loudspeaker directivity characteristics (as defined in the introduction) in a room with a constant $T_{60}$ time as a function of frequency. The choice of such a transfer function represents a loudspeaker that introduces no spectral inconsistencies and a room that has little influence on the spectrum at the listening position. The resulting transfer function has an amplitude that decreases with frequency, reflecting omnidirectional directivity at low frequencies, exciting the reverberant sound field more, and more directional at higher frequencies. Chosen for the presented experiment is a target response with a \SI{3}{\dB} drop over the frequency range \SI{20}{\Hz} to \SI{20}{\kHz}. As a result of this target function, the DRR will still be frequency-dependent, with a smoother transition compared to a traditional loudspeaker placed in a room with uncontrolled acoustical properties.

\section{Perceptual Evaluation}\label{sec:PE} 
\begin{figure*}
    \begin{center}
        \includegraphics[width = \linewidth]{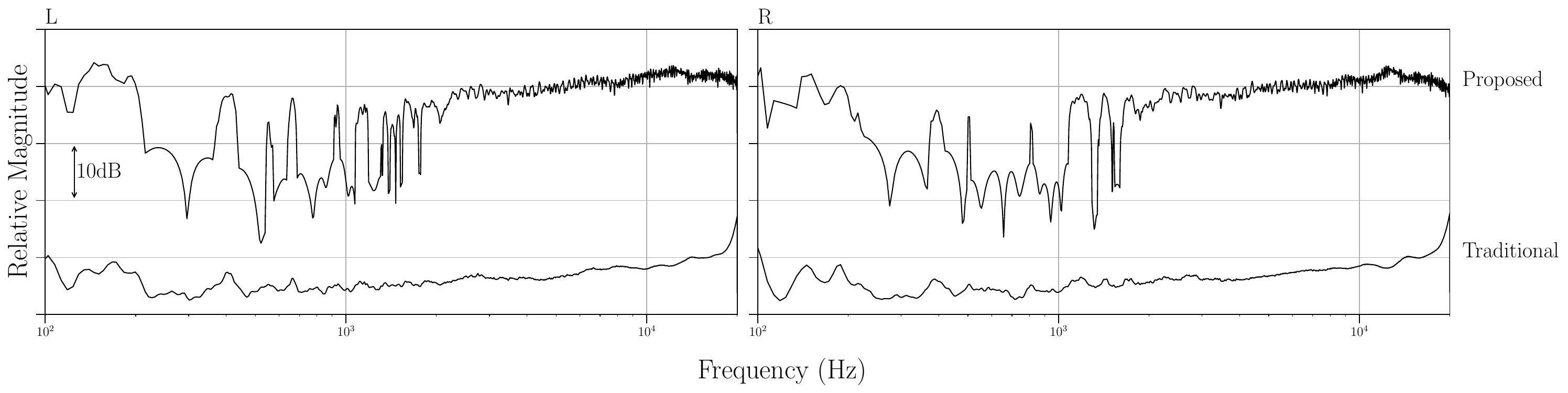}
        \caption{\label{fig:filter_responses} The magnitude responses of the filters used in the presented perceptual evaluation, excluding the commercial algorithm filters, are shown in the left and right figures. The left figure (L) displays the filters for the left channel, with the top showing the filter from the supporting loudspeaker filter in the proposed approach, \eqref{eq:ProposedComp}, the bottom showing the inverse filter applied to the primary loudspeaker, \eqref{eq:TraditionalComp}. The right figure is the same for the right channel (R).}
    \end{center}
\end{figure*}

\subsection{Experimental Setup}
\begin{figure}
    \begin{center}
        \includegraphics{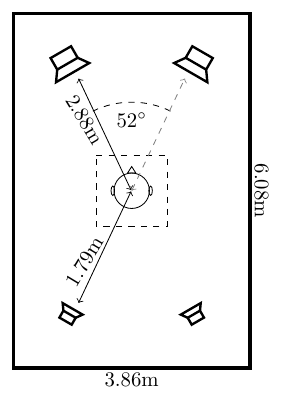}
        \caption{\label{fig:PEDiagram}A diagram of the room used for the presented perceptual evaluation. Including two primary loudspeakers each with a corresponding supporting loudspeaker. The listener is seated within an acoustically transparent curtain, with an additional curtain between the seating position and the door.}
    \end{center}
\end{figure}

In order to perceptually evaluate the proposed method, a stereo pair of loudspeakers was used, where each loudspeaker is independently compensated by a supporting loudspeaker. A stereo setup was chosen to represent a typical triangular loudspeaker setup commonly found in the home, and to reproduce more spatial imaging than a mono loudspeaker setup. Two Bowers \& Wilkins D3 loudspeakers formed our primary loudspeaker pair, powered by a 150W amplifier, and connected to a MacBook running the experiment software in Max 7 with a Behringer DAC. Each primary loudspeaker was supported by a Genelec 8030 studio monitor, connected to test software using the same DAC. Figure \ref{fig:PEDiagram} presents the experiment room dimensions and the layout of the loudspeakers. The room used for the presented experiment had $T_{60}\text{'s}$  that conformed with the IEC 268-13 standard (an average of \SI{0.4}{\s}), but deviated significantly from an optimal listening environment that the target function is based on. Also included in Figure \ref{fig:PEDiagram} is the acoustically transparent curtain that, along with an upward-firing lamp, ensured participants had no knowledge of the number and position of the loudspeakers within the room. Combined with a pathway to the central container, subjects were also not aware of the shape, size, and orientation of the room.

\subsection{Audio Stimuli}
The playback methods alongside the proposed approach were standard uncompensated stereo playback, a simple amplitude based inverse filter similar to that proposed by \cite{bouchard2006inverse}, and a well established commercial room compensation algorithm. Responses to all filters used in the perceptual evaluation (excluding the commercial filter, which is not available), are shown in Figure \ref{fig:filter_responses}.

The included amplitude based inverse filter was calculated by inverting the averaged energy, and 1/3 octave smoothed amplitude response of the two microphones in the listening area, with appropriate regularisation, to reach the defined target function, 
\begin{equation}\label{eq:TraditionalComp}
    w(\omega) = \frac{\overline{H(\omega)} D(\omega)}{|H(\omega)|^2 + \beta (\omega) |H(\omega)|} ,
\end{equation}
where $\overline{H(\omega)}$ represents the complex conjugate and $\beta(\omega)$ is a frequency-dependent regularisation factor. Regularisation both ensures the filter only compensates for the response within a specified frequency range, and ensures that the filter does not attempt to compensate for nulls in the transfer function that would result in ringing in the resulting filter. The $\beta$ was chosen to compensate only within the frequency range of \SI{70}{\Hz} and \SI{20}{\kHz}. Outside this range, $\beta$ is set to 1 and 0.001 within. The filter was transformed to the time domain using an IFFT and subsequently converted to minimum phase, to ensure the time domain response of the filter was not cyclic \cite{smithfilters}. The same target function used for the proposed approach is also used for the  traditional inverse filter, with the omission of the target function constraints.

Subjects rated algorithms based on three pieces of music, indicated in Table \ref{tab:Content}. All stimuli were 1 minute long, subjects were able to listen for as long as required and could seamlessly switch between conditions with a short cross-fade time. The content has been selected to include three different genres, and one of the pieces is a live recording in order to make the content selection as broad as possible, Table \ref{tab:Content} contains information on the content. The ISO 532-2 loudness model \cite{ISO_532-2_2017} was used to match all stimuli to a reference stimuli of 72dBA, which was found to be a suitable listening level by the authors, followed by a loudness matching verification by a tonmeister with loudness matching experience.

\subsection{Experimental Design}
A simultaneous preference-based rating method has been used to evaluate the performance of the proposed method against several comparison conditions. Subjects were asked to rate the presented algorithms based on preference using a bank of continuous sliders repeated across several pieces of music, and were prompted to use the full range of the sliders. As there is no reference in this experiment, preference was expected to be compared to an internal reference. The presentation order of the music was randomised; the preference rating slider order was consistent across all subjects.

After subjects concluded their ratings, a series of questions were asked about their listening experience. The questions were whether they had any comments on what they heard, if there are of normal hearing, if they consider themselves to be an expert listener, and how many loudspeakers they perceived throughout the experiment and where the loudspeakers were located.

\subsection{Assessors}
Eight self declared normal hearing subjects participated in the presented experiment, all working in the field of acoustics and all but two regularly participating in loudspeaker critical listening training sessions.

\begin{table*}
    \caption{\label{tab:Content}Details on the pieces of music used for the described perceptual evaluation}
        \begin{tabular}{l l l c} 
            \hline
            \hline
            \textbf{Title} & \textbf{Artist/Album} & \textbf{Genre} & \textbf{Time Stamp}  \\ 
            \hline
            Limehouse Blues & Jazz at the Pawnshop  & Jazz& (2:30 - 3:30) \\
            Thinking Out Loud & Ed Sheeran & Pop & (1:00 - 2:00) \\
            Prologue: La selva  & Orfeo Chaman & Classical & (1:00 - 2:00) \\
            \hline
            \hline
        \end{tabular}
\end{table*}

\subsection{Results}\label{sec:PERes}
\begin{figure}
    \begin{center}
        \includegraphics{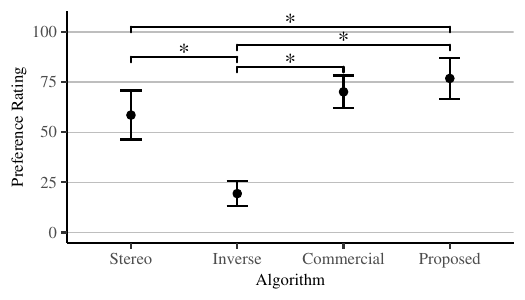}
        \caption{\label{fig:results} Preference Results from the described perceptual evaluation. Significance is indicated with an `*', for an alpha level of 0.05, using a pairwise t-test with a Holm-Bonferroni correction.}
    \end{center}
\end{figure}

Figure \ref{fig:results} presents the results from the subjective evaluation, including mean values, 95\% confidence intervals, and significance indicators ($*$). After confirming normality, a two-way repeated measures analysis of variance was conducted with main effects of content and reproduction method, as well as their interaction. This analysis found that content did not have a significant effect on preference ratings, nor did the interaction between content and reproduction method. However, the reproduction method did have a significant effect; therefore, results across all content and participants will be used throughout this section to analyse the effect of the reproduction method. In order to analyse the consistency of ratings within subjects and reproduction methods, the coefficient of variation (CV) ($\frac{\sigma}{\mu}$, where $\sigma$ is the standard deviation and $\mu$ the mean) has been calculated. The results from this analysis show that $5/8$ subjects displayed low variance between content ratings, with a mean CV of $18.9\%$, but $3/8$ displayed a higher variation of $>60\%$, with an overall mean CV of $38.2\%$. Given that there is no reference condition for the presented experiment, and subjects are making ratings based upon some internal reference, the low CV values suggest that for the majority of the subjects, there is some consistency to their ratings, which coincides with no statistical effect of content. Confidence intervals were calculated using the standard t-distribution formula accounting for sample size and variability. Significance was tested using pairwise t-tests with a Holm-Bonferroni correction, reported along side Cohen's d, and statistical power ($1-\beta$) values.

The research question for the presented perceptual evaluation is to investigate whether the perception, both spatial and timbral, of a primary loudspeaker can be altered by a supporting compensation loudspeaker, without the supporting loudspeaker being spatially perceived as an additional source. Whilst subjects were asked to rate methods based on preference, a difference in preference can be interpreted as a perceivable difference. However, if no difference in preference is observed, this does not mean that there is no perceivable difference, as ratings may have been influenced by several contradicting internal criteria. Therefore, only pairs with significant differences can be used to draw conclusions about perceivable differences.

The combination of the significant difference between uncompensated stereo reproduction and the proposed approach preference ratings ($p < 0.05, d = 0.69, (1-\beta) = 0.64$), alongside the answers to the post-rating questions (results presented in Table \ref{tab:questionnaire}) where no subject with prior knowledge about the proposed approach perceived any sources to the rear of the listening position, indicates the proposed approach works as intended. For the question regarding perceiving sources anywhere but the primary source, subjects were asked to describe the loudspeaker configuration and positioning, the classification criterion for the presented results was that subjects indicated all sources to be in the frontal hemisphere.

A second comparison can be made between the proposed approach, uncompensated stereo, and the commercial room compensation algorithm. There is no significant difference between the proposed and commercial approach ($p = 0.303, d = 0.31, (1-\beta) = 0.18$), prohibiting us from making a formal conclusion as to the preference between these two approaches. However, whilst a significant difference between uncompensated playback and the proposed approach is observed ($p < 0.05, d = 0.69, (1-\beta) = 0.64$), the same significance is not seen between uncompensated playback and the commercial room compensation algorithm ($p = 0.154, d = 0.47, (1-\beta) = 0.35$).

By comparing the inverse filter to other comparison stimuli, it is clear that such a filter has a negative effect on preference. Indicated by the significant difference between it and uncompensated playback ($p < 0.001, d = 1.7, (1-\beta) = 0.99$).

\begin{table}[ht]
    \caption{\label{tab:questionnaire}Summary of responses to the post-rating questionnaire (N = 8).}
    \begin{ruledtabular}
        \begin{tabular}{lc}
            \textbf{Question} & \textbf{Count} \\
            \hline
            Self-declared normal hearing & 8 \\
            Self-declared expert listener & 8 \\
            Regularly partakes in listening training & 6 \\
            Perceived sources from anywhere but primary sources & 0 \\
        \end{tabular}
    \end{ruledtabular}
\end{table}

\section{Technical Evaluation}\label{sec:ObjMes}
Technical measures are often used to evaluate room compensation algorithms; see \cite{cecchi_room_2017} for an overview of some of these methods. Whilst for other methods of room compensation technical measures may be a useful tool for comparison, due to the nature of the proposed approach, the presented technical measures do not correspond with the previously discussed perceptual evaluation.

\begin{figure*}
    \begin{center}
        \includegraphics[width=\linewidth]{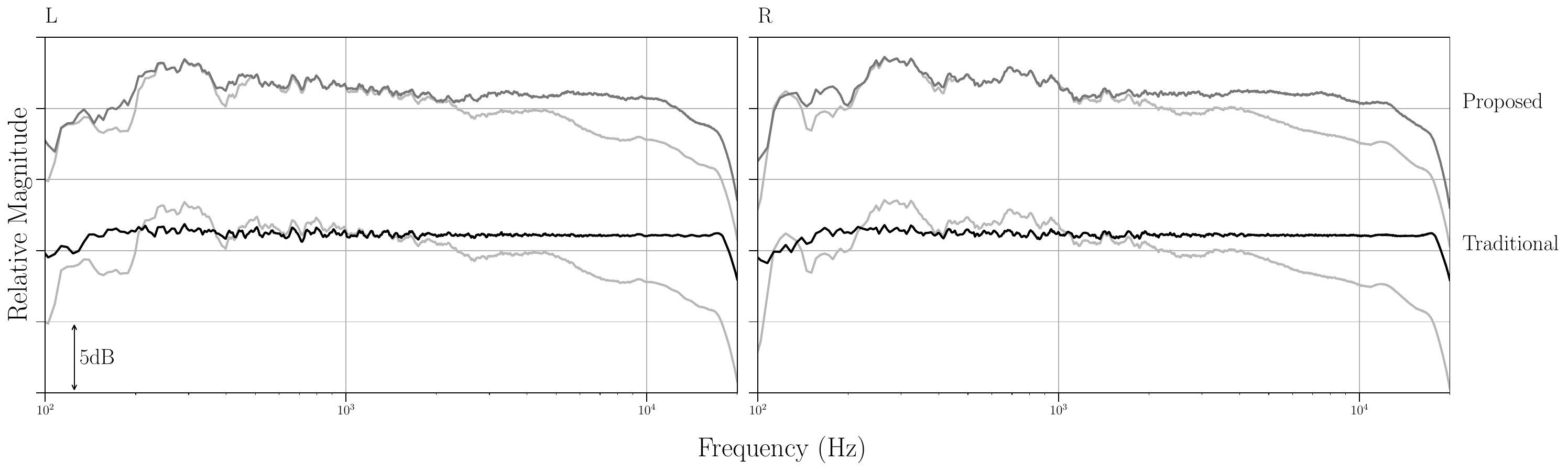}
        \caption{\label{fig:ResultingResponse}Frequency plots representing the transfer function before (light grey) and after compensation (black for traditional and dark grey for proposed) for the traditional inverse method and the proposed method. The left figure shows the responses for the left channel and the right figure shows the corresponding compensations for the right channel.}
    \end{center}
\end{figure*}

Figure \ref{fig:ResultingResponse} presents frequency responses of the LRTFs from the room used for the perceptual evaluation, along with the resulting transfer functions after applying a direct inverse filter and the proposed approach. From these plots, the limitations of the proposed method can be seen, where there are peaks remaining in the transfer function, as energy emitted by the primary loudspeaker can not be subtracted, and, in addition there are notches to ensure the precedence effect is not broken.

To technically compare the two approaches, the spectral deviation of the resulting transfer functions is compared across both channels. Spectral deviation determines the extent to which energy deviates from the average energy as a function of frequency,
\begin{equation}\label{specdev1}
    S_D = \sqrt{\frac{1}{Q_h-Q_l-1}\sum_{i=Q_l}^{Q_h}(20\log_{10}(\lvert Y(\omega_i)\rvert) - D)^2} \; , 
\end{equation}
\vspace*{-0.6cm}
where
\begin{equation}\label{specdev2}
    D = \frac{1}{Q_h-Q_l-1}\sum_{i=Q_l}^{Q_h}20\log_{10}\lvert Y(\omega_i)\rvert \;\;.
\end{equation}

The frequency range of evaluation spans from $Q1 = \SI{100}{\Hz}$ to $Q2 = \SI{20}{\kHz}$, accounting for the low-frequency roll-off of the loudspeaker, of which compensation is not attempted. For spectral deviation measures in this work, transfer functions will be smoothed by a 1/3 octave window to better represent the spectral fidelity of the human auditory system. If the target amplitude of the compensation algorithm is set to be constant as a function of frequency, the spectral deviation measure assesses the technical effectiveness of amplitude based compensation.

Table \ref{tab:SpecDevRes} presents spectral deviation measures for the two compensation methods applied to the left and right channels. These results indicate an averaged \SI{3.5}{\dB} difference between the proposed and traditional compensation methods, in favour of the traditional method of room compensation. This difference suggests that at any frequency, the deviation may exceed the just noticeable difference thresholds for spectral sensitivity \cite{moore2003introduction}.

\begin{table}[ht]
    \caption{\label{tab:SpecDevRes}Spectral deviation results for left and right channels for the two compared methods. A lower value indicates a more technically effective compensation method.}
    \begin{ruledtabular}
        \begin{tabular}{l c c} 
            \textbf{Algorithm} & \multicolumn{2}{l}{\textbf{Spectral Deviation $S_D$(dB)}}\\ 
            \hline
            & Left channel & Right channel\\
            Traditional: \eqref{eq:TraditionalComp}&  1.1  & 1.1  \\
            Proposed:  \eqref{eq:ProposedComp}& 4.5 & 4.7  \\
        \end{tabular}
    \end{ruledtabular}
\end{table}

\section{Direct to Reverberant Ratio Analysis} \label{sec:DRRAnalysis}
One of the claimed benefits of the proposed method discussed in Section \ref{sec:Theory} is the ability to control the Direct to Reverberant Ratio (DRR) as a function of frequency, in addition to the spectrum of the sound field. Section \ref{sec:Theory} demonstrates that traditional room compensation methods alter both the direct and reverberant sound fields by applying a filter directly to the source. In contrast, in the proposed method the direct sound is left unaltered only adding energy to the reverberant sound field. To demonstrate this effect in practice, both traditional and proposed compensation methods were applied to impulse responses generated by a room simulation algorithm, and the DRR was analysed before and after compensation.

The room simulation software RAZR \cite{wendt_computationally-efficient_2014} independently generates the direct sound, early reflections, and late reflections. This allows both methods to be independently applied to the direct and reverberant components (the reverberant component consists of the early and late reflections), such that the DRR can be accurately calculated. The simulated room for this purpose has the same dimensions as that used in the presented perceptual evaluation, and has the source and receiver positions for the left channel. Within RAZR, a spherical head model is used to model the directivity of a traditional loudspeaker as proposed in \cite{brown_structural_1998}. The left channel of a dichotic microphone pair is used to generate impulse responses, and the DRR is calculated by looking at the ratio of energy between the direct and reverberant sound field as a function of frequency, as in \eqref{eq:DRR} for the uncompensated sound field. 

The room compensation methods used in this analysis are identical to those used in the previous section, technically evaluating the two approaches. Where a spectrally flat target function is selected to allow a simple analysis of the resulting sound field. For the Proposed method of compensation, the DRR is calculated by the ratio between the direct sound of the primary loudspeaker, and the reverberant sound that is the reverberant sound of the primary loudspeaker and the total response of the supporting loudspeaker, as in \eqref{eq:PropOptDRR}. For the traditional approach, the DRR can be calculated with \eqref{eq:TradOptDRR}.

\begin{figure}
    \begin{center}
        \includegraphics[width = \linewidth]{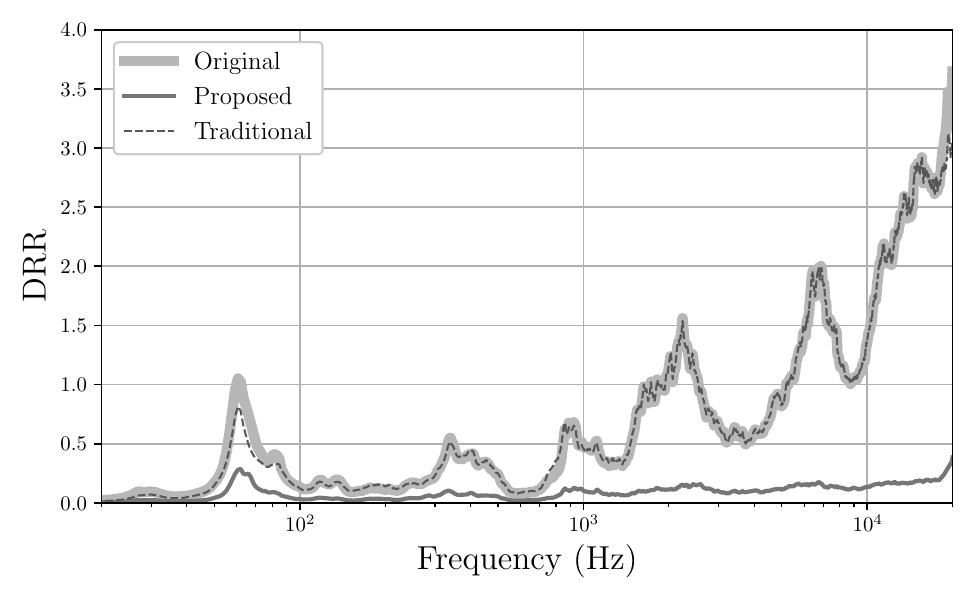}
        \caption{\label{fig:DRRAnalysis} A comparison of the DRR for an uncompensated (thick light grey), proposed compensated (dark grey), and traditionally compensated (dashed black) sound field. Responses are generated in RAZR and the DRR is smoothed with 1/3 octave windows.}
    \end{center}
\end{figure}

Figure \ref{fig:DRRAnalysis} shows the DRR as a function of frequency for the sound field before compensation, after traditional room compensation, and the proposed method of room compensation. The spectrum of all sound fields has been smoothed on a 1/3 octave basis to align more with the sensitivity of the human auditory system, and to allow for ease of analysis. It is clear from this figure that the uncompensated DRR (Thick light grey line) significantly changes as a function of frequency. As would be expected, as frequency increases, the DRR increases as there is less energy emitted into the reverberant sound field due to traditional source directivity. Such a DRR, depicting traditional uncompensated reproduction, may cause shifts in distance perception across the frequency range, where peaks in the DRR may be audible and sound unnatural. The DRR of the traditionally compensated sound field (dashed black line) does not differ significantly from the uncompensated DRR, due to the inability to control the direct and reverberant sound fields independently (as discussed in Section \ref{sec:Theory}). In contrast, the proposed method of room compensation is able to alter the reverberant sound field independently of the direct sound, which can be seen in the DRR for the proposed approach (thin dark grey line). Due to the spectrally flat target function, the DRR is also significantly more frequency-independent than both the uncompensated and traditionally compensated sound fields. The overall DRR is also much lower than both of the comparison sound fields, as energy has been added to the reverberant sound field, decreasing the total DRR across the frequency range.

\section{General Discussion}\label{sec:GeneralDiscussion}
The proposed method of room compensation aims to compensate for spectral and spatial inaccuracies in the loudspeaker room transfer function. In order to achieve room- and loudspeaker-independent sound reproduction, energy is added to the perceived reverberant sound field. As discussed in Section \ref{sec:Theory}, traditional compensation methods are limited to compensating for the spectral response of the LRTF. By independently altering the perceived reverberant sound field, the DRR can be altered, compensating for spectral and spatial inaccuracies introduced by the loudspeaker room combination.

To verify the proposed method of room compensation, a subjective evaluation has been conducted, investigating whether the perception, both spatial and timbral, of a primary loudspeaker can be altered by a supporting compensation loudspeaker, without the supporting loudspeaker being spatially perceived as an additional source. Results presented in Section \ref{sec:PE} show that preference ratings are significantly higher for the  proposed compensation compared to uncompensated playback, indicating a difference in perception when the supporting source is present. Supplemented by a post-rating questionnaire, no subject perceived an additional source. This evidence suggests that the proposed approach is able to alter the perception of the loudspeaker without the listener spatially perceiving an additional source.

From the results discussed in Section \ref{sec:PERes}, the traditional method of room compensation significantly decreased subjects preference ratings. This is likely because steps were not taken to limit the compensation filter, in order not to introduce perceptual artefacts to the reproduction experience. These results are consistent with those from \cite{Norcross2004SubjectiveIO}, where inverse filtering has also been shown to negatively impact reproduction quality.

In Section \ref{sec:ObjMes}, technical measures are presented for the proposed method and the traditional room compensation approach, showing that the traditional method of room compensation outperforms the proposed approach. This is partially due to the inability to remove energy from the reverberant sound field and the limitations introduced to ensure the precedence effect is not broken. However, the technical evaluation of the proposed approach is brought into question when compared to the subjective evaluation of the two compensation methods. In Section \ref{sec:PE}, a significant difference in favour of the proposed approach is observed comparing the same two room compensation methods. This difference could be due to perceptual artefacts introduced by the traditional room compensation method that are absent in the proposed approach,the increased capability of the proposed approach to compensate for spatial reproduction accuracy, or a combined effect of both factors.

The DRR of the uncompensated sound field has been compared to a traditional method of room compensation and the proposed method. This analysis shows that the proposed method is able to control the DRR of a sound field, whereas traditional methods are not. The proposed approach, therefore, is likely to alter the spatial and distance perception of the reproduced content, giving a more consistent or smoother perception of distance to the listener as a function of frequency. It can also be noted that the DRR for the proposed approach is lower across the entire frequency range, which is likely to increase the distance perceived by the listener. It is interesting to note that traditionally, rooms with a lower $T_{60}$ times, therefore a higher DRR, are preferred \cite{kaplanis2019perception}. However, in the presented study, the proposed approach, with a lower DRR, was preferred over uncompensated playback, with a higher DRR. This could be due to the well controlled nature of the reverberant sound field for the proposed approach, which may not be consistent with the reverberant sound field of our listening environments.

Given the significant difference in preference ratings between the commercial compensation algorithm and the traditional inverse approach, it is likely that the traditional inverse approach introduced audible artefacts not present in the commercial compensation algorithm. As discussed in the introduction, many methods of reducing or avoiding perceptual artefacts introduced in the process of room compensation have been proposed. However, the same methods to reduce artefacts in the traditional inverse filter are used in the proposed approach, without artefacts being introduced. For the proposed compensation method, the nature of altering only the perceived reverberant sound field naturally evades introducing artefacts that are common in traditional inverse filters. This effect can be explained by the inherently stochastic properties of the reverberant sound field, which results in robustness to additional artefacts introduced by the supporting loudspeaker.

The length of traditional room compensation filters has been discussed by \citet{fielder2003analysis}, suggesting that the filter length should not exceed the temporal masking threshold of the human auditory system \cite{moore2003introduction}. However, for short filter lengths, the filter's ability to compensate for the playback room is limited. In contrast to traditional methods of room compensation, the filter is applied to the perceived reverberant sound field, meaning temporal masking in the direct sound no longer plays a role in the perception of the filter, as the direct sound is left unaltered. This allows the filter to be much longer in length without introducing perceptual artefacts.

In conventional compensation techniques, where modifications are made only to the primary source, care must be taken to ensure the phase properties within and between sources preserve temporal compactness and interaural phase differences. In the proposed method, the direct component of the primary source is left unaltered avoiding these problems while still being able to compensate for the reverberant sound field.

In the proposed method, a supporting loudspeaker is used to compensate for the reverberant sound of a primary loudspeaker. An idea has been explored in early experimental stages where the supporting channel (the delayed spectrally modified version of the content) is also reproduced by the primary loudspeaker. Whilst this has not been formally investigated, the authors found it to produce unnatural spatial effects on the reproduced sound field. A potential explanation for this is due to the fixed fluctuation of a coincident primary and supporting source, differing from a non-coincident primary and supporting source positions where such fixed fluctuation patterns are not created. In the latter case, source LRTFs will vary with the listener's head movements, creating a variable addition of the two signals, physically relating more to natural mono loudspeaker playback, where both the direct and reverberant components vary independently with head movement. This may be important as it has been suggested that the perception of a sound field is represented more as an average across different positions, rather than the perception of a single position \cite{de2001spatial}.

The results in Section \ref{sec:ObjMes} show that the traditional method of room compensation outperforms the proposed approach objectively. Figure \ref{fig:ResultingResponse} shows the frequency regions where the proposed approach falls short of the traditional approach. One area where the proposed approach does not objectively perform well is at higher frequencies, where limitations introduced to the target function reduce the compensation accuracy, compared to the traditional approach where these limitations are not required. However, at low frequencies, where target function limitations do not have such a great effect on the target function (as seen in Figure \ref{fig:targetlimitations}), the proposed approach still does not perform to the same level as the traditional approach. This is likely due to the larger wavelength at lower frequencies, where the summation of the primary and supporting sources may be subject to unforeseen phase interactions that are not accounted for in the filter design or by the supporting loudspeaker delay and decorrelation.

The results of the discussed experiment suggest that the accuracy of the DRR may be important for sound reproduction. This is consistent with research investigating methods for dedicated acoustic recording and playback, where the direct and reverberant sound fields are separately reproduced and optimised independently, controlling the DRR, with the aim of accurately reproducing the recorded content \cite{Grosse2015}.

\section{Conclusion}\label{sec:Conclusion}
A method of room compensation has been proposed, whereby energy is added to the perceived reverberant sound field to compensate for spectral and spatial inaccuracies in loudspeaker reproduction. This is achieved by delaying and modifying the transfer function of a secondary supporting loudspeaker, adding energy to the room in a frequency-selective manner without the listener perceiving the supporting loudspeaker.

It has been shown theoretically how the proposed method is able to not only alter the spectrum of the LRTF, but also modify the DRR, improving upon traditional techniques where only the spectrum can be altered. A perceptual evaluation demonstrates that the proposed approach performs as expected, altering and improving the perception of a primary source (Timbral and Spatial) without the listener perceiving an additional supporting source. The presented evaluation also shows that in the specific circumstances of our experiment, the proposed approach performs comparably, if not exceeding, the performance of an established commercial algorithm. Technical measurements have been shown not to correlate with preference-based responses in the perceptual evaluation, due to the relevance of auditory phenomenon that are not taken into account in technical measures. An analysis of the proposed approach's impact on DRR, in simulations, supports the presented theoretical motivation.

\begin{acknowledgments}
This project has received funding from the European Union's Horizon 2020 research and innovation programme under the Marie Skłodowska-Curie grant agreement No 956369. The authors thank Stephan Töpken for helpful comments on an earlier draft of the manuscript.
\end{acknowledgments}

\section*{Author Declarations}
The authors declare no conflict of interest.

\section*{Ethics Approval}
The authors have obtained formal informed consent from all the listening test participants involved in the study.

\section*{Data Availability}
The data that support the findings of this study are available from the corresponding author upon reasonable request.

\bibliography{References.bib}
\end{document}